\documentclass[12pt]{article}
 \usepackage{graphicx,epsfig}% Include figure files
 \usepackage{amssymb}

\begin{document}
 \begin{center}
 {\Large Recoil studies in the reaction of $^{12}C$ ions \,\,\,\, \\ with the enriched
isotope $^{118}Sn$.}

\vspace{8mm}
 A.R.Balabekyan \footnote{Yerevan State University, Armenia}$^{,*}$,
 A.S.Danagulyan$^{1}$,
 J.R.Drnoyan$^{1,3}$,
 N.A.Demekhina \footnote{Yerevan Physics Institute},
 G.H.Hovhannisyan$^1$,\\
J.Adam \footnote{JINR, Dubna, Russia}$^{,}$ \footnote{INF AS
R\v{e}z, Czech Republic}, V.G.Kalinnikov$^3$, M.I.Krivopustov$^3$,
 V.S.Pronskikh$^3$, V.I.Stegailov$^3$,\\
 A.A.Solnyshkin$^3$, V.M.Tsoupko-Sitnikov$^3$\\
S. G. Mashnik \footnote[5]{\textsf{Los Alamos National Laboratory,
Los Alamos,USA}}$^)$,
 K. K. Gudima \footnote [6]{\textsf{Institute of Applied Physics, Academy of Science of
 Moldova, Chi\c{s}in\u{a}u}}$^)$ \\

 \end{center}
 \begin{center}
 {\bf Abstract}
 \end{center}

The recoil properties of the product nuclei from the interaction
of 2.2 GeV/nucleon $^{12}C$ ions %the LHE JINR
from Nuclotron  of the Laboratory of High Energies (LHE), Joint
Institute for Nuclear Research (JINR) at Dubna with a $^{118}$Sn
target have been  studied using catcher foils.

The experimental data were analyzed using the mathematical formalism
of the standard two-step vector model. The results for $^{12}C$ ions
are compared with those for deuterons and protons.

Three different Los Alamos versions of the Quark-Gluon String Model
(LAQGSM)were used for comparison with our experimental data.

\section{Introduction}

Studies of the interactions of high-energy projectiles with complex
nuclear targets lead to an overall picture of great complexity. The
violent processes that are associated with a high multiplicity of
emitted particles and lead to products a hundred or more masses,
removed from the target are not well understood.

On the other hand, if one investigates less violent encounters,
which can be classified under the generic term of target
fragmentation some simplifications will be obtained . A target
fragment is a major remnant of the portion of target nucleus which
appears to have been the spectator to the region of
projectile-target interaction. Linear momentum transfer between a
projectile and a target fragment is expected to be low. For
heavy-ion projectiles, an equivalent process in the moving system
results in projectile fragments which have nearly the beam velocity.
Many observations of target or projectile fragmentation appear
consistent with two hypotheses originally associated with peripheral
hadron-hadron interactions at high energies \cite{Bog}. The
hypothesis of {\it limiting fragmentation} predicts that both cross
sections and energy spectra of fragments should become independent
of bombarding energy at high energies. The second, factorization,
asserts that these fragment projectiles can be written as a product
of target and projectile factor. Both cross-sections of formation of
residual nuclei and their recoil properties can be used for
investigation of hypothesis mentioned above.

The recoil properties of nuclei are determined often via
"thick-target thick-catcher" experiment by using induced activation
method . In such experiments, the thicknesses of the target and
catcher foils are larger than the longest recoil range. The
quantities measured are the fractions $F$ and $B$ of product nuclei
that recoil out of the target foil into the forward and backward
directions, respectively.

The results of the experiment are usually proceeded by the standard
two-step vector representation \cite{sugar}--\cite{win2}.

The aim of our investigation is to obtain the kinematic
characteristics of residuals in reactions of $^{12}C$ ions with
enriched tin isotope $^{118}$Sn. Comparison of the results obtained
in $^{12}C$ and proton and deuteron induced reactions will be used
to examine the extent of the deviations from factorization exhibited
by these properties.

\section{Experimental setup and results}

Targets of enriched tin isotope $^{118}$Sn were irradiated at the
Nuclotron  of the
%Laboratory of High Energies (
LHE,
%), Joint Institute for Nuclear Research (
JINR
%) at Dubna
by $^{12}C$ ion beams with energies of 2.2 GeV/nucleon. Irradiations
were of 10 hours and had round form with a diameter of 2 cm. For
beam monitoring, we used the reaction $^{27}$Al$(^{12}C,X)^{24}$Na
with cross sections of $14.2\pm0.2$ mb \cite{dam}.

The target consisted of a high-purity target metal foil of size
20x20 mm$^2$ sandwiched exactly by one pair of Mylar foils of the
same size, which collected the recoil nuclei in the forward or
backward directions with respect to the beam. The enrichment of the
target was 98.7 \%, the thickness of each target foil was 66.7
$mg/cm^2$ and the number of target piles was 11. The whole stack,
together with an Al beam-monitor foil was mounted on a target holder
and irradiated in air.

After irradiation the target foils and all of the forward and
backward catcher foils from one target pile were collected
separately, and assayed for radioactivities nondestructively with
high-purity Ge detectors at LNP, JINR for one year. The radioactive
nuclei were identified by characteristic $\gamma$ lines and by their
half-lives. The spectra were evaluated with the code package
DEIMOS32 \cite{frana}.

The kinematic characteristics of around twenty product nuclei were
obtained. The relative quantities of the forward- and
backward-emitted nuclei  (relative to the beam direction) were
calculated from relations:
\begin{equation}
F=N_F/(N_t+N_F+N_B); \,\,\,\,  B=N_B/(N_t+N_F+N_B)
\end{equation}
% Eq. (4)
where $N_F$ , $N_B$ , $N_t$ are the numbers of nuclei emitted in
forward and backward catchers and formed in target foils,
respectively. The recoil parameters obtained in these experiments
are the forward-to-backward ratio, $F/B$, and the mean range,
$2W(F+B)$. [The mean range of the recoils is somewhat smaller than
$2W(F+B)$, but it is conventional to refer to the latter quantity as
range]. The mathematical formalism of the standard two-step vector
model \cite{sugar}-\cite{win1} was used to proceed the experimental
results.

The following assumptions are made in this model:

\hspace*{10mm}(1) In the first step, the incident particle interacts
with the target nucleus to form an excited nucleus with velocity
$v$, momentum $P$, and excitation energy $E^*$.

\hspace*{10mm}(2) In the second step, the excited nucleus loses mass
and excitation energy to form the final recoiling nucleus with an
additional velocity $V$, which in general will have a distribution
of values and directions.

The results of the recoil experiments depend on the range-energy
relation of the recoiling nuclei. It is convenient to express this
relation as  \cite{win1}:
\begin{equation}
R=kV^n ,
\end{equation}

The parameters  $k$ and $n$ in this formalism are obtained by
fitting the range dependence on energy of accelerated ions within
the region from 0.025 to 5 MeV/nucleon \cite{range}. It is possible
to calculate $\eta=v/V$ and $v$ in frame of two-step vector model
\cite{win1}, knowing the $F/B$ ratio of the experiment.

Our experimental results are shown in Tables 1. We note that
uncertainties concerning definite quantities in our tables are not
listed to keep the tables concise. These uncertainties are about
15--20\%. As shown in Fig.\ 1, the ratios $F/B$ for $^{12}C$ ions
induced reactions are of the order of $\sim 3 - 4$ for heavy product
nuclei and decrease to about $\sim 1.5$ for light residuals. Such
dependence could be explained by different mechanisms for the
production of nuclei in different mass regions. Heavy nuclei are
produced mainly via the spallation mechanism, with its products more
in the forward direction, while light residual nuclei may be
produced by multifragmentation or fission-like processes that lead
to an isotropic distribution in the frame of excited residual
nuclei. For understanding the mechanism of formation of residual
nuclei a comparison with theoretical calculation was made.

Theoretical calculations by the LAQGSM03.01 \cite{LAQGSM0301},
LAQGSM03.S1 \cite{LAQGSM03S1}, and LAQGSM03.G1 \cite{LAQGSM03S1}
models with our experimental results are compared. As can be seen
from Fig.\ 1, there is some disagreement between experimental data
and theoretical results by all three versions of LAQGSM considered
here. Theoretical calculations well describe experimental ones only
for the products in mass region $24<A<70$. It can be possible to
explain it as follows: when making such a comparison, we first
recognize that the experiment and the calculations differ in that:
(i) the experimental data were extracted assuming the "two-step
vector model" \cite{win1,win2}, while the LAQGSM calculations were
done without the assumptions of this model; and (ii) the
measurements were performed on foils (thick targets), while the
calculations were done for interactions of $^{12}C$ and $^{118}Sn$
nuclei (thin targets).

LAQGSM03.01 \cite{LAQGSM0301} is the latest modification of the Los
Alamos version of the Quark-Gluon String Model \cite{LAQGSM}, which
in its turn is an improvement of the Quark-Gluon String Model
\cite{QGSM}. It describes reactions induced by both particles and
nuclei as a three-stage process: IntraNuclear Cascade (INC),
followed by preequilibrium emission of particles during the
equilibration of the excited residual nuclei formed during the INC,
followed by evaporation of particles from or fission of the compound
nuclei.

LAQGSM03.S1 \cite{LAQGSM03S1} is exactly the same as LAQGSM03.01,
but considers also multifragmentation of excited nuclei produced
after the preequilibrium stage of reactions, when their excitation
energy is above 2A MeV, using the Statistical Multifragmentation
Model (SMM) by Botvina et al.\ \cite{SMM} (the ``S" in the extension
of LAQGSM03.S1 stands for SMM).

LAQGSM03.G1 \cite{LAQGSM03S1} is exactly the same as LAQGSM03.01,
but uses the fission-like binary-decay model GEMINI of Charity et
al.\ \cite{GEMINI}, which considers evaporation of all possible
fragments, instead of using the GEM2 model \cite{GEM2} (the ``G"
stands for GEMINI).

Many details and further references on LAQGSM03.01 and its 03.S1 and
03.G1 versions may be found in \cite{mashnik}.

Fig.\ 2 shows the dependence of the fragment kinetic energy,
$T_{kin}$, on the fractional mass loss $\Delta A/A$. The comparison
with theoretical calculations by the LAQGSM03.01 \cite{LAQGSM0301},
LAQGSM03.S1 \cite{LAQGSM03S1}, and LAQGSM03.G1 \cite{LAQGSM03S1}
models shows that for mass region of residuals $24<A<117$ the
experimental results are described better by the LAQGSM03.G1 model.
As one can see in article \cite{bon} formation of residual nuclei in
mass region  $A_t/3<A<A_t/2$ can be explained by fragmentation as
well as fission-like processes. In our previous publication
\cite{bal} we described the formation of medium-weight nuclei
($7<A<60$) by multifragmentation mechanism while heavy residual
nuclei were formed by evaporation mechanism. The fission-like binary
decay model of GEMINI used by LAQGSM03.G1 can be regarded somehow in
the middle between multifragmentation and conventional evaporation,
and it looks like it describes better the present experimental data
in a wide mass region of the products. Owing to the proportionality
between longitudinal momentum and excitation energy of the remnant
in spallation regime \cite{bron} the mean velocity along the beam
direction of the remnant $v_{II}$ should increase linearly with
$\Delta A/A_t$ ($\Delta A = A_{targ}-A_{res}$, where $A_{targ}$ is
the mass number of target and $A_{res}$ is the mass number of
product nuclei) in this regime. It is interesting to note that the
measured forward velocity $v_{II}$ increases practically linearly
with the increase of $\Delta A/A_t$ , but remains approximately
(roughly) constant at around $\Delta A=60$ (Fig.\ 3). In our past
publications \cite{bal1,bal} for proton- and deuteron-induced
reactions we revealed that such behavior of dependence is
conditioned by change of mechanism of production of residuals. The
same picture was obtained for the production of the measured nuclei
by $^{12}C$ ions.

Fig. 4 shows the dependence of the mean ranges of fragments on the
targets mass numbers. Our results for the target $^{118}Sn$ conform
well with the results for the targets in mass range from $Cu$ to $U$
at the $^{12}C$ ions beam energy 18.5 GeV \cite{cole}. The ranges
increase monotonically with target $A_t$ and for a given target vary
in an essentially inverse manner with product $A$.

In our recent paper \cite{bal}, we have investigate the recoil
properties of the product nuclei on the enriched $^{118}Sn$ target
at the 3.65 GeV protons and 7.3 GeV deuterons beams. Therefore it is
interesting to determine the applicability of factorization to the
present data.

Fig. 5 shows a comparison of $F/B$ ratios from 26.4 GeV  $^{12}C$
ions and 3.65 GeV protons bombardment of $^{118}Sn$ and $^{12}C$
ions and 7.3  GeV deuterons. For all three projectiles the behaviors
of ratio $F/B$ via $\Delta A = A_{targ}-A_{res}$ are similar:
decrease monotonically with increasing $\Delta A$. The $F/B$ ratios
for $^{12}C$ ions are slightly lower than those obtained with 3.65
GeV protons and the same average value than obtained with 7.3 GeV
deuterons. The weighted average of the ratios of $F/B$ for
individual products are $0.77 \pm 0.16$ and $1.03 \pm 0.20$
respectively.

Fig. 6 shows the product mass dependence of ratios of $v_{II}$ from
26.4 GeV $^{12}C$ ions and 3.65 GeV protons bombardment of
$^{118}Sn$  and $^{12}C$ ions and 7.3 GeV deuterons. The weighted
average of the ratios of individual ranges being $0.84 \pm 0.25$ for
$^{12}C$ and protons and $1.11 \pm 0.49$ for $^{12}C$ and deuterons.
In general, our comparison of recoil properties and cross-sections
(partially unpublished) of residual nuclei of the interaction of
$^{118}Sn$ with 26.4 GeV $^{12}C$ ions with similar data for
comparably energy (per nucleon) protons and deuterons indicates that
the production cross-sections and recoil properties attain the
regime of limiting fragmentation. The comparison of $v_{II}$ for
$^{12}C$ ions and protons and deuterons support a factorization
within errors for the energy of projectiles $>2 GeV/nucleon$.

\section{Conclusion}
The recoil properties of the product nuclei from the interaction of
2.2 GeV/nucleon $^{12}C$ ions with a $^{118}$Sn target were
obtained. The dependence of recoil properties via fractional mass
loss shown the change of production mechanism for medium light
residuals ($A<60$). A comparison with the same properties for 3.65
GeV protons and 3.65 GeV/nucleon deuterons was made. Limiting
fragmentation and factorization in this energy region were obtained.

\section*{Acknowledgments}

The authors would like to express their gratitude to the operating
personnel of the JINR Nuclotron and Synchrophasotron for providing
good beam parameters. This work was supported partially by the US
DOE.

\vspace*{2cm}
%\newpage
 \begin{center}
 \textbf{Table 1}. Kinematic characteristics of product nuclei on
 $^{12}C$ ions induced reactions.

 \vspace*{5mm}
 \begin{tabular}{|c|c|c|c|c|c|}  \hline
 %\label{tabl-1}
 Product    &  F/B  & $\eta$& 2W(F+B) &  T$_{kin.}$ (MeV)  & v$(MeV/amu)^{1/2}$ \\ \hline
 $^{24}Na$  & 1.40  & 0.084 &  4.82$\pm$ 1.02 & 20.24$\pm$ 4.29  & 0.1089  \\ \hline
 $^{28}Mg$  & 1.54  & 0.107 &  4.41$\pm$ 0.94 & 19.18$\pm$ 4.07  & 0.1255  \\ \hline
 $^{43}K$   &  1.55 & 0.108 &  2.86$\pm$ 0.61 & 15.11$\pm$ 3.20  & 0.0909  \\ \hline
 $^{44}Sc$  &  2.32 & 0.207 &  2.49$\pm$ 0.53 & 12.21$\pm$ 2.59  & 0.1543  \\ \hline
 $^{52}Mn$  &  1.88 & 0.157 &  3.27$\pm$ 0.69 & 21.51$\pm$ 4.56  & 0.1426  \\ \hline
 $^{67}Gd$  &  1.02 & 0.005 &  1.55$\pm$ 0.33 & 7.17$\pm$  1.52  & 0.0024  \\ \hline
 $^{71}As$  & 2.48  & 0.223 &  1.50$\pm$ 0.32 &  7.12$\pm$ 1.51  & 0.1001  \\ \hline
 $^{73}Se$  &  3.33 & 0.292 & 1.51$\pm$ 0.32  & 7.33$\pm$ 1.55   & 0.1309  \\ \hline
 $^{86m}Y$  &  3.38 & 0.296 &  1.05$\pm$ 0.22 & 4.37$\pm$ 0.93   & 0.0943  \\ \hline
 $^{77}Br$  & 2.35  & 0.210 &  1.35$\pm$ 0.29 & 6.11$\pm$ 1.29   & 0.0837  \\ \hline
 $^{89}Zr$  & 2.87  & 0.258 & 0.73$\pm$0.15   & 2.45$\pm$ 0.52   & 0.0605  \\ \hline
 $^{90}Mo$  &  2.32 & 0.207 &  1.28$\pm$ 0.27 & 6.35$\pm$ 1.35   & 0.0778  \\ \hline
 $^{90}Nb$  &  3.49 & 0.303 &  1.01$\pm$ 0.21 & 4.08$\pm$0.86    & 0.0912  \\ \hline
 $^{93m}Mo$ & 4.33  & 0.351 &  0.92$\pm$  0.19& 3.69$\pm$ 0.78   & 0.0988  \\ \hline
 $^{93}Tc$  &  1.63 & 0.121 &  1.26$\pm$ 0.27 & 6.23$\pm$ 1.32   & 0.0445  \\ \hline
 $^{94}Tc$  & 3.14  & 0.278 &  0.98$\pm$0.21  & 4.21$\pm$ 0.89   & 0.0833  \\ \hline
 $^{95}Tc$  & 3.64  & 0.312 &  0.61$\pm$ 0.13 & 1.97$\pm$ 0.42   & 0.0635  \\ \hline
 $^{97}Ru$  & 4.02  & 0.334 &  0.59$\pm$ 0.13 & 1.93$\pm$ 0.41   & 0.0667  \\ \hline
 $^{99m}Rh$ &  3.16 & 0.280 & 0.50$\pm$ 0.11  & 1.48$\pm$ 0.31   & 0.0484  \\ \hline
 $^{109}In$ & 4.47  & 0.358 & 0.17$\pm$ 0.03  & 0.28$\pm$ 0.06   & 0.0256  \\ \hline
 $^{110}Sn$ &  2.35 & 0.211 &0.039$\pm$ 0.008 & 0.029$\pm$ 0.006 & 0.0049  \\ \hline
 $^{111}In$ &  3.34 & 0.293 & 0.13$\pm$  0.03 & 0.19$\pm$ 0.04   & 0.0169  \\ \hline
 $^{117m}Sn$&  1.08 & 0.020 &  0.17$\pm$ 0.03 & 0.27$\pm$ 0.06   & 0.0014  \\ \hline
 \end{tabular}
 \end{center}

\newpage

\vspace*{10mm}
% \newpage
 \begin{figure}[h]
 %\hspace*{-7mm}
 \includegraphics[scale=0.5]{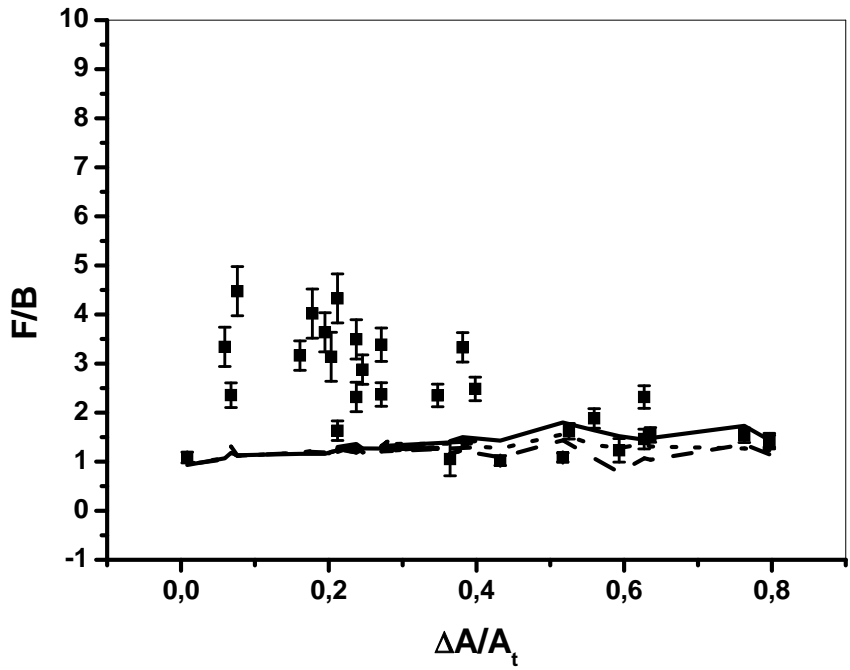}
  \vspace*{-40mm}
 \caption{
$F/B$ versus the fractional mass losses $\Delta A/A_t$: line shows
the calculation by LAQGSM.01, dash line shows the calculation by
LAQGSM.S1, dot line shows the calculation by LAQGSM.G1.
  }
 \end{figure}

\vspace*{35mm}
 %\newpage
 \begin{figure}[h]
 %\hspace*{-7mm}
 \includegraphics[scale=0.5]{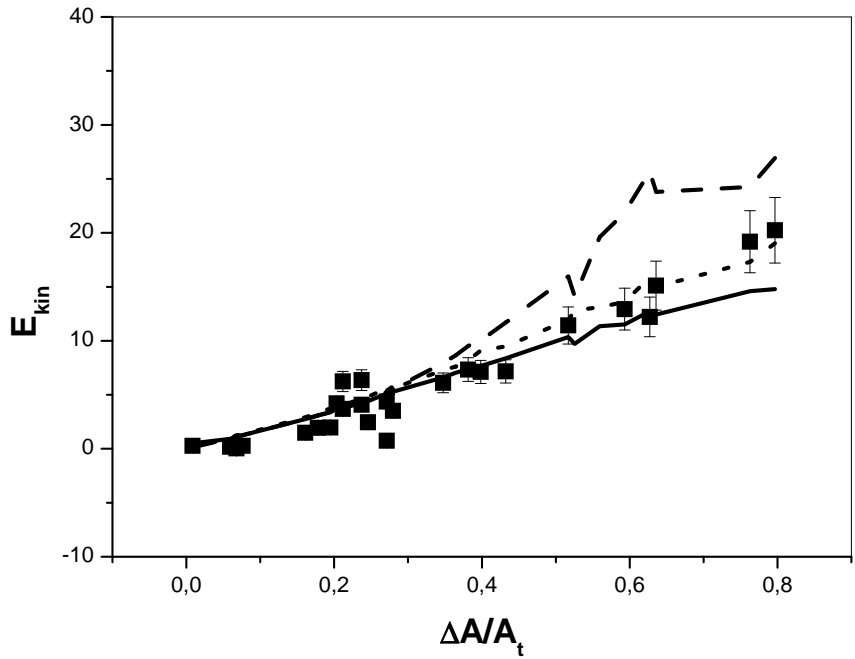}
  \vspace*{-40mm}
 \caption{
 Dependence of the kinetic energy of the product nuclei
on the fractional mass losses $\Delta A/A_t$:  dot line shows the
calculation by LAQGSM.G1, line shows the calculation by LAQGSM.01,
dash line shows the calculation by LAQGSM.S1.
  }
 \end{figure}

 \newpage

\vspace*{20mm}
  %\newpage
 \begin{figure}[h]
 %\hspace*{-7mm}
 \includegraphics[scale=0.5]{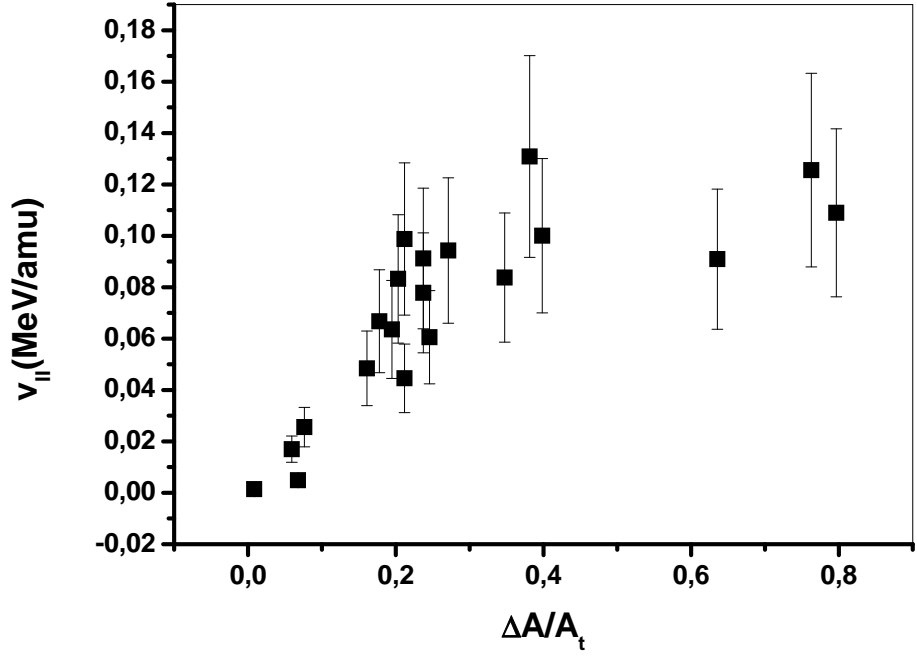}
  \vspace*{-35mm}
 \caption{
 Dependence of the forward velocity of excited nuclei
on the fractional mass losses $\Delta A/A_t$:
  }
 \end{figure}

\vspace*{30mm}

 \begin{figure}[h]
 %\hspace*{-7mm}
 \includegraphics[scale=0.5]{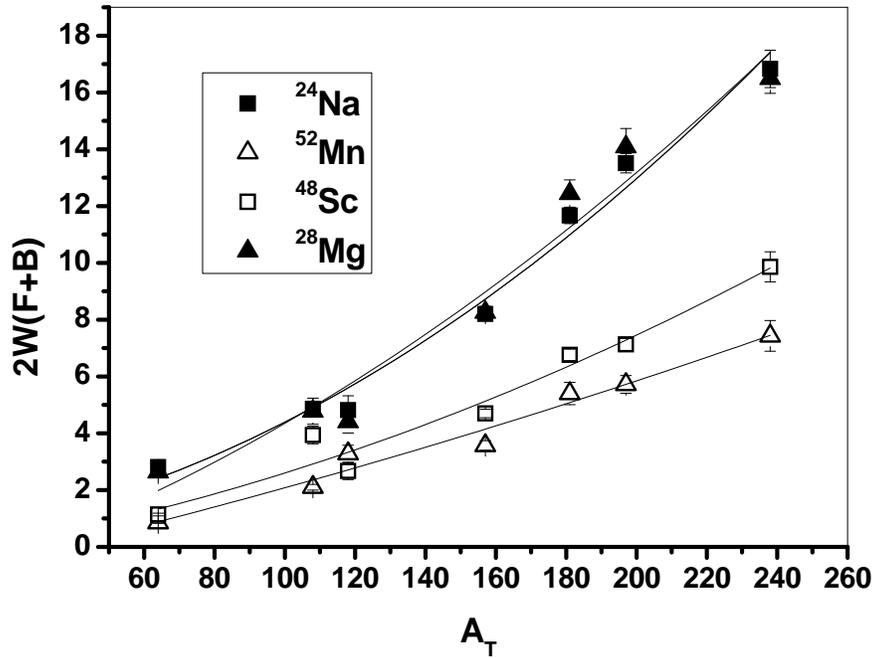}
  \vspace*{-35mm}
 \caption{
Mean ranges of fragments emitted in the interaction of $^{12}C$ ions
with targets of mass $A_t$: Basic data for 18.5 GeV $^{12}C$ ions
were taken from \cite{cole}. Our data are only for $A_T$=118. The
curves show the trends in the data.
  }
 \end{figure}

 \newpage

\vspace*{10mm}
 \begin{figure}[h]
 \hspace*{-7mm}
 \includegraphics[scale=0.35]{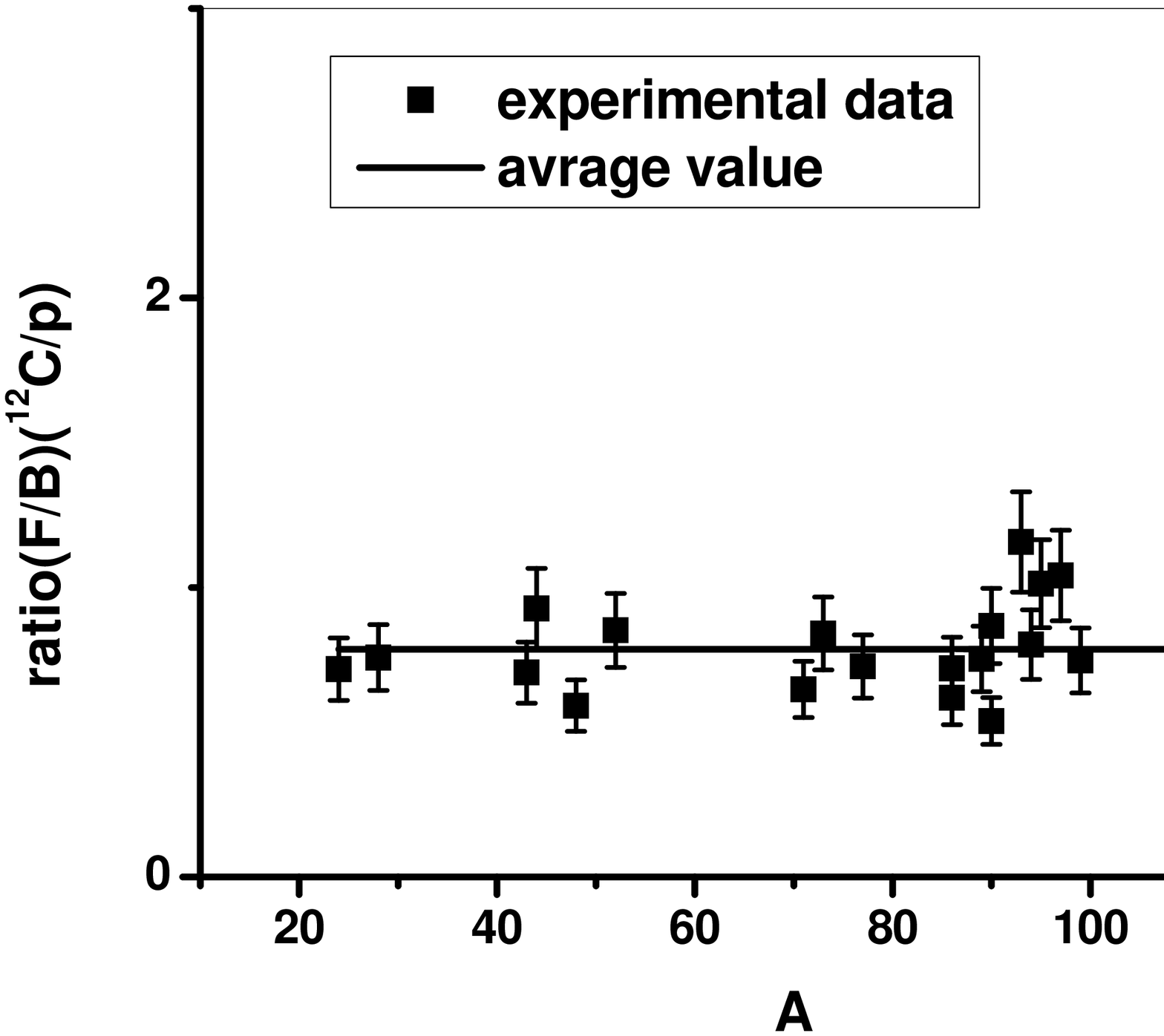}
 \hspace*{-15mm}
 \includegraphics[scale=0.35]{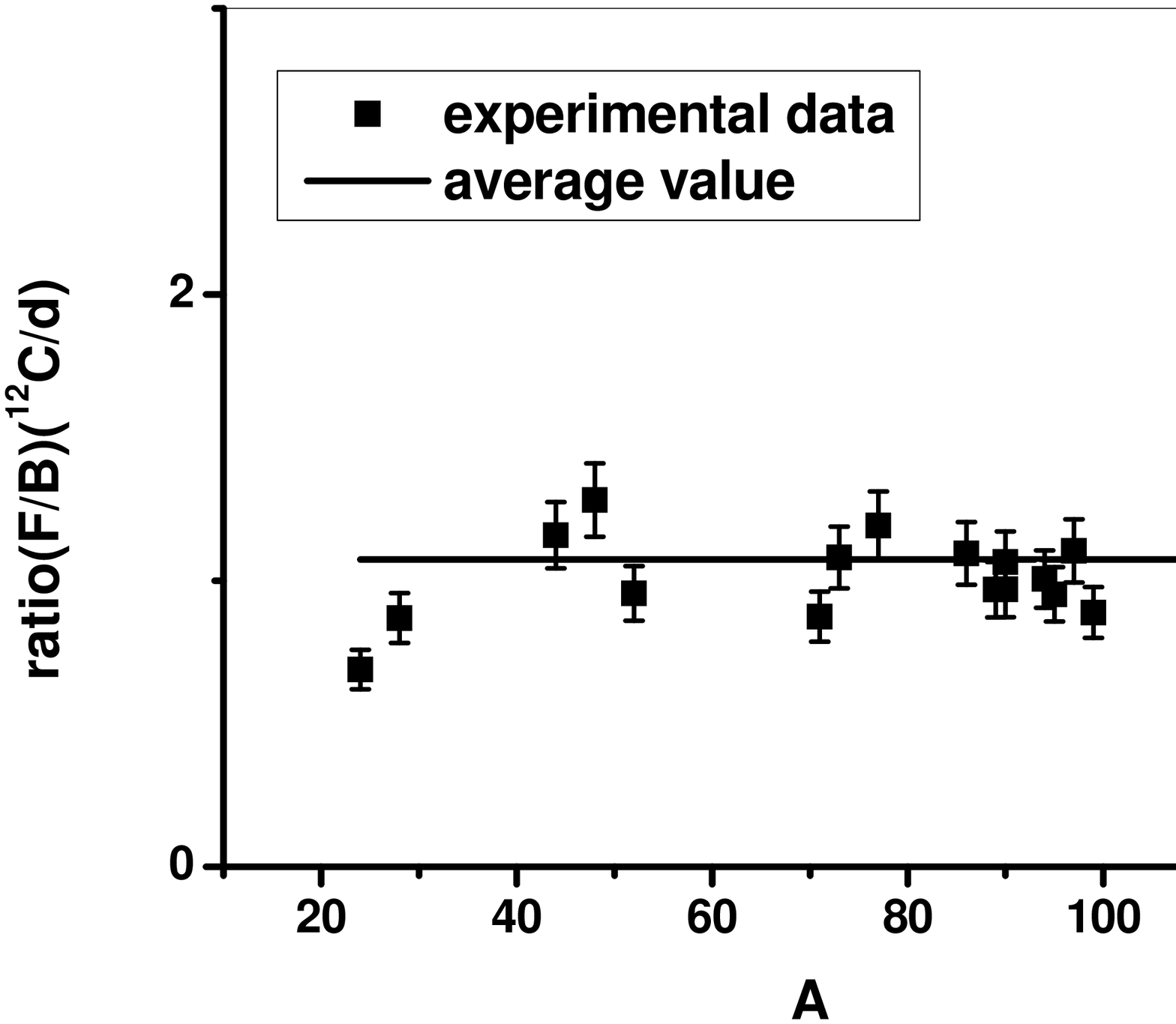}
 \vspace*{-35mm}
 \caption{Product mass dependence of ratios of F/B from 26.4 GeV
 $^{12}C$ ions and 3.65 GeV protons bombardment of $^{118}Sn$
 and $^{12}C$ ions and 7.3  GeV deuterons.
  }
 \end{figure}

 \vspace*{20mm}
 \begin{figure}[h]
 \hspace*{-7mm}
 \includegraphics[scale=0.35]{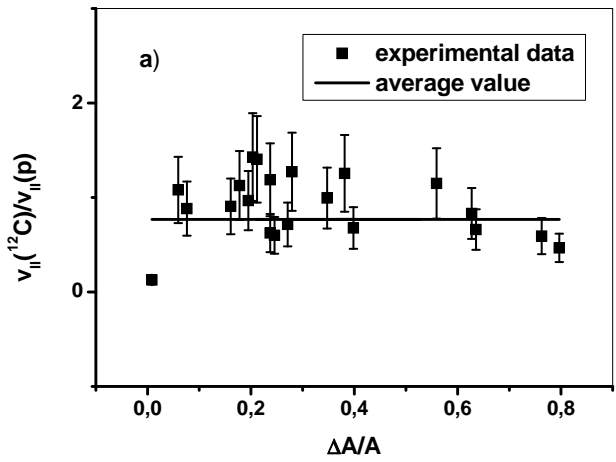}
 \hspace*{-15mm}
 \includegraphics[scale=0.35]{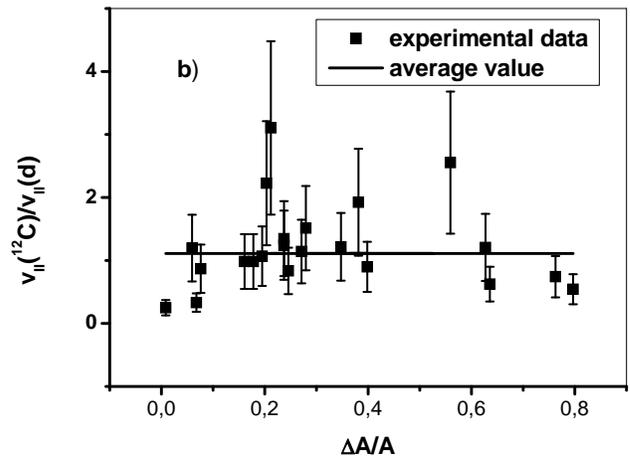}
 \vspace*{-35mm}
 \caption{Product mass dependence of ratios of $v_{II}$ from 26.4 GeV
 $^{12}C$ ions and 3.65 GeV protons bombardment of $^{118}Sn$
 and $^{12}C$ ions and 7.3  GeV deuterons.
  }
 \end{figure}


\newpage
 \begin{thebibliography}{99}

\bibitem{Bog}
H. Boggild and T. Ferbel
%"Inclusive reactions"
Annual Reviews Nucl.\ Sci.\ 24 (1974) 451-514.

\bibitem{sugar}                                                         %[1]
N. Sugarman, M. Campos and K. Wielgoz,
%``Recoil Studies of High-Energy Proton Reaction in Bismuth",
Phys.\ Rev.\ 101 (1956) 388.
%No. 1, pp. 388-397

\bibitem{win1}                                                        %[2]
L. Winsberg,
%``The analysis of thick-target thick-catcher nuclear recoil experiment"
Nucl.\ Instrum.\ Methods 150 (1978) 465.
%No. 3, pp. 465-477

\bibitem{win2}                                                         %[3]
L. Winsberg,
%``Systematics of particle-nucleus reactions. I. Parameters,"
Phys.\ Rev.\ C 22 (1980) 2116.
% 2116-2122 (1980) [Issue 5 - November 1980 ]

\bibitem{dam}
O. Dandinsuren, V. M. Dyachenko et al., Preprint JINR P1-87-932.
(1987).

\bibitem{frana}                                                       %[11]
J. Fr\'{a}na,
%``Program DEIMOS32 for gamma-ray spectra evaluation,"
J.\ Radioanal.\ Nucl.\ Chem. 257 (2003) 583.
% Journal of Radioanalytical and Nuclear Chemistry, Volume: 257, Issue: 3
%September 2003 pp. 583-587

\bibitem{kort}                                                          %[4]
R. G. Kortelling, C. R. Toren and E. K. Hyde,
%``Energy Spectra of Fragments from Silver and Uranium Bombarded
%with 5.0-GeV Protons,"
Phys.\ Rev.\ C 7 (1973) 1611.
%1611-1626 (1973)[Issue 4 - April 1973 ]

\bibitem{range}                                                       %[12]
L. C. Northcliffe and R. F. Schilling,
%``Range and Stopping Power Tables for Heavy Ions,"
Nucl.\ Data A 7 (1970) 233.
%No. 1-2, pp. 233-463

\bibitem{data}                                                          %[5]
L. Winsberg,
%``Ranges of projectiles with 8 =< Z =< 20 in Ti, Fe, Ni, Cu, Ag, and Au
%for energies of 0.0125 to 12.0 MeV/nucleon,"
At.\ Data Nucl.\ Data Tables 20 (1977) 389.
% Volume: 20, Issue: 5 November, 1977 pp. 389-396

\bibitem{LAQGSM0301}                                                    %[17]
S. G. Mashnik, K. K. Gudima, M. I. Baznat, A. J. Sierk, R. A. Prael
and N. V. Mokhov,
%``CEM03.01 and LAQGSM03.01 Versions of the
%Improved Cascade-Exciton Model (CEM)
%and Los Alamos Quark-Gluon String Model (LAQGSM) Codes,"
LANL Report, LA-UR-05-2686, Los-Alamos (2005).

\bibitem{LAQGSM03S1}                                                    %[18]
S. G. Mashnik, K. K. Gudima, M. I. Baznat, A. J. Sierk, R. A. Prael
and N. V. Mokhov,
%``CEM03.S1, CEM03.G1, LAQGSM03.S1, and LAQGSM03.G1
%Versions of CEM03.01 and LAQGSM03.01 Event-Generators,"
LANL Report, LA-UR-06-1764, Los-Alamos (2006).

\bibitem{LAQGSM}                                                       %[19]
K. K. Gudima, S. G. Mashnik and A. J. Sierk,
%``User Manual for the code LAQGSM,"
LANL Report LA-UR-01-6804, Los Alamos (2001),
http://lib-www.lanl.gov/la-pubs/00818645.pdf.

\bibitem{SMM}                                                           %[25]
J. P. Bondorf, A. S. Botvina, A. S. Iljinov, I. N. Mishustin and K.
Sneppen,
%``Statistical Multifragmentation of Nuclei,"
Phys.\ Rep.\ 257 (1995) 133.%--221. %No. 3

\bibitem{GEMINI}                                                        %[26]
R. J. Charity, M. A. McMahan, G. J. Wozniak, R. J. McDonald, L. G.
Moretto, D. G. Sarantites, L. G. Sobotka, G. Guarino, A. Pantaleo,
L. Fiore, A. Gobbi and  K. D. Hildenbrand,
%``Systematics of Complex Fragment Emission in Niobium-Induced Reactions,"
Nucl.\ Phys.\ A  483 (1988) 371. %--405. %No.2

\bibitem{GEM2}                                                          %[24]
S. Furihata,
%``Statistical Analysis of Light Fragment Production from Medium Energy
%Proton-Induced Reactions,''
Nucl.\ Instrum.\ Methods B 171 (2000) 252; %--258;
PhD thesis, Tohoku University, March, 2003.

\bibitem{QGSM}                                                           %[20]
N. S. Amelin, K. K. Gudima and V. D. Toneev,
%``The Quark-Gluon String Model and Ultrarelativistic Heavy-Ion Collisions,"
Yad.\ Fiz.\ 51 (1990) 512 %--523.
[Sov.\ J. Nucl.\ Phys. 51 (1990) 327]; %--333
%``Ultrarelativistic Nucleus-Nucleus Collisions in a Dynamical Model of
%Independent Quark-Gluon Strings,"
Yad.\ Fiz.\ 51 (1990) 1730 %--1743.
[Sov.\ J. Nucl.\ Phys. 51 (1990) 1093]; %--1101
%``Further Development of the Model of Quark-Gluon Strings for the
%Description of High-Energy Collisions with a Target Nucleus,"
Yad.\ Fiz.\ 52 (1990) 272 %--282
[Sov.\ J. Nucl.\ Phys. 52 (1990) 172]; %--178
N. Amelin,
%``Physics and Algorithms of the Hadronic Monte-Carlo Event Generators.
%Notes for a Developer,"
CERN/IT/ASD Report CERN/IT/99/6, Geneva, Switzerland (1999);
%and JINR/LHE, Dubna,
%Russia;
%``GEANT4, Users's Documents, Physics Reference Manual," last update
%08/04/99:
http://wwwinfo.cern.ch/asd/geant4/G4UsersDocuments/UsersGuides/
PhysicsReferenceManual/html/PhysicsReferenceManual.html.


\bibitem{mashnik}
 S. G. Mashnik, K. K. Gudima, R. E. Prael, A. J. Sierk, M. I.
Baznat, and N. V. Mokhov, "CEM03.03 and LAQGSM03.03 Event Generators
for the MCNP6, MCNPX, and MARS15 Transport Codes" Invited lectures
presented at the Joint ICTP-IAEA Advanced Workshop on Model Codes
for Spallation Reactions, February 4-8, 2008, ICTP, Trieste, Italy,
LA-UR-08-2931, Los Alamos (2008); E-print: arXiv:0805.0751v2
[nucl-th]; IAEA Report INDC(NDS)-0530, Distr. SC, Vienna, Austria,
August 2008, p. 51.

\bibitem{bon}
J. Bondorf, R. Donangelo, I. N. Mishustin and H. Schulz
%Statistical multifragmentation of nuclei
Nucl.Phys. A 444 (1985) 460-476.

\bibitem{bron}
M. Bronikowski and N. T. Porile
%Recoil properties of target fragments from the interaction of silver
%with 218 GeV $^{16}O$ ions
Phys. Rev. C45 (1992) 1389-1391.

\bibitem{bal1}                                                       %[15]
V. Aleksandryan, J. Adam, A. Balabekyan, A. S. Danagulyan, V. G.
Kalinnikov, G. Musulmanbekov, V. K. Rodionov, V. I. Stegailov,
% and
J. Frana,
%``Formation of Residual Nuclei with Medium Mass Number in the Reaction of
% Protons with Separated Tin Isotopes,"
Nucl.\ Phys.\ A 674 (2000) 539.
%No 3-4, pp. 539-549

\bibitem{bal}
A. R. Balabekyan, A. S. Danagulyan, J. Drnoyan et al.
%Recoil Products from $p+^{118}Sn$ and $d+^{118}Sn$ at 3.65 GeV/A.
Yad. Fiz.,70, (2007),1940, Physics of Atomic Nuclei, 70, (2007),
1889.

\bibitem{cole}
G. D. Cole and N. T. Porile
%Recoil properties of fragments emitted in the interaction of
%complex nuclei with relativistic $^{12}C$ ions and protons.
Phys. Rev. C25 (1982) 244-255.


\end{thebibliography}
\end{document}